\documentclass[aip,apl,reprint]{revtex4-1}

\usepackage{graphicx}

\begin{document}

\title{Attracting shallow donors: Hydrogen passivation in (Al,Ga,In)-doped ZnO} 


\author{Masahiko Matsubara}
\thanks{M. Matsubara and M. N. Amini contributed equally to this work.}
\author{Mozhgan N. Amini}
\thanks{M. Matsubara and M. N. Amini contributed equally to this work.}
\author{Rolando Saniz}
\author{Dirk Lamoen}
\author{Bart Partoens}
\affiliation{CMT \& EMAT, Departement Fysica, Universiteit Antwerpen, Groenenborgerlaan 171, B-2020 Antwerpen, Belgium}


\date{\today}

\begin{abstract}
The hydrogen interstitial and the substitutional $\mathrm{Al}_{\mathrm{Zn}}$, $\mathrm{Ga}_{\mathrm{Zn}}$ and $\mathrm{In}_{\mathrm{Zn}}$ are all shallow donors in ZnO and lead to $n$-type conductivity. Although shallow donors are expected to repel each other, we show by first principles calculations that in ZnO these shallow donor impurities attract and form a complex, leading to a donor level deep in the band gap. This puts a limit on the $n$-type conductivity of (Al,Ga,In)-doped ZnO in the presence of hydrogen.
\end{abstract}


\maketitle


The incorporation of hydrogen in a material can strongly affect its electronic properties. Furthermore, its behavior depends on the host into which it is introduced. Mostly, interstitial hydrogen is amphoteric, i.e. it can act either as a donor ($\mathrm{H_i}^+$) or an acceptor ($\mathrm{H_i}^-$)~\cite{vandewalle2003}. Which of these two prevails depends on the Fermi level: it acts as a donor in $p$-type materials and as an acceptor in $n$-type materials. Consequently, it counteracts the conductivity caused by extrinsic dopants. It is important to realize that this passivation of extrinsic dopants by hydrogen is not caused by the formation of a neutral complex. The formation of such a neutral complex may just be the consequence after compensation, i.e. after the trapping of a free hole or electron by hydrogen. Afterwards, the positively charged H and negatively charged acceptor, or negatively charged H and positively charged donor attract and may form a neutral complex.

In ZnO however, the behavior of hydrogen is not amphoteric~\cite{vandewalle2003}. Nominally undoped ZnO shows $n$-type conductivity and it was first predicted by first-principles calculations based on density functional theory~\cite{vandewalle00}, and later experimentally verified~\cite{cox01,*hofmann02}, that interstitial hydrogen acts as a shallow donor in ZnO and can cause this $n$-type behavior.

ZnO is a wide band gap ($\sim$3.4 eV at room temperature~\cite{chen98}) semiconductor with many possibilities for technological applications as a transparent conductor in solar cells~\cite{nuruddin01}, flat panel displays~\cite{gordon00}, etc. It is well-known that the electronic conductivity of ZnO can be dramatically improved by doping it with group III elements Al, Ga or In~\cite{minami00}. This makes ZnO an alternative for indium tin oxide, which shows highest conductivity among transparent conducting oxides but is less abundant and therefore more expensive. It is then also important to know what limits the conductivity of ZnO doped by Al, Ga or In. Properties such as formation energy, energy band structure and equilibrium geometry of Al, Ga and In doped ZnO were studied systematically and it was shown that all three dopants are indeed shallow donors and that they possess the intrinsic qualities to be good $n$-type transparent conducting oxides~\cite{xu11}.

As the concentration becomes high, the interaction between defects becomes unavoidable. Because it is very difficult to remove hydrogen from the crystal growth environment, it is then also natural to ask what the influence is of interstitial hydrogen on Al, Ga or In doped ZnO. Recently a number of experiments were reported where both H and group III elements are co-doped into ZnO. It was shown that the conductivity of ZnO co-doped with H and group III elements is better than that doped with group III elements up to certain amount of H inclusion~\cite{liu07,*lee08,*lee09,*tark09,*kimdh10}.

Apart from single hydrogen dopants, the influence of hydrogen on doped ZnO structures has been studied before by first principles calculations. It was found that hydrogen passivates nitrogen dopants~\cite{xiaonan05}, silver dopants~\cite{he2009} and Zn vacancies~\cite{karazhanov2009} in ZnO, and forms complexes. This is not very surprising, as in all these cases, hydrogen, which is a (positive) shallow donor, is attracted to a (negative) acceptor impurity. This is not unique to hydrogen.  Also donor-like As$_{\mathrm{Zn}}$ and acceptor-like V$_{\mathrm{Zn}}$ attract each other and form a complex in As doped ZnO~\cite{limpijumnong04}. Recently, also the attractive interaction between two donor impurities was predicted by first principles studies in ZnO, namely the deep donor V$_{\mathrm{O}}$ (the oxygen vacancy) and the shallow donor ${\mathrm{Zn}_i}$~\cite{kim09} (the Zn interstitial). In this Letter, we consider the two shallow donors interstitial hydrogen $\mathrm{H}_i$ and substitutional $\mathrm{X}_{\mathrm{Zn}}$ with $\mathrm{X}=\mathrm{Al}$, $\mathrm{Ga}$ or $\mathrm{In}$.


We performed density functional theory calculations as implemented in the Vienna ab initio simulation package (VASP) code~\cite{kresse96,kresse99}.
In order to examine interactions between defects, a large supercell consisting of 192 atoms was considered~\cite{Note1}.
A single Al, Ga or In atom was chosen to substitute a Zn atom, while for the H atom, an extensive number of positions around the group III element were considered. For all these configurations the formation energies for different charge states were calculated using the method of Ref.~\cite{janotti07}. This method was recently confirmed to give reliable results for the transition energies between different charge states of the impurity, both for ZnO with interstitial H~\cite{janotti11} as for ZnO doped with a group III element~\cite{xu11}, by comparing the results with hybrid functional calculations.

\begin{figure}
\includegraphics[width=8.5cm]{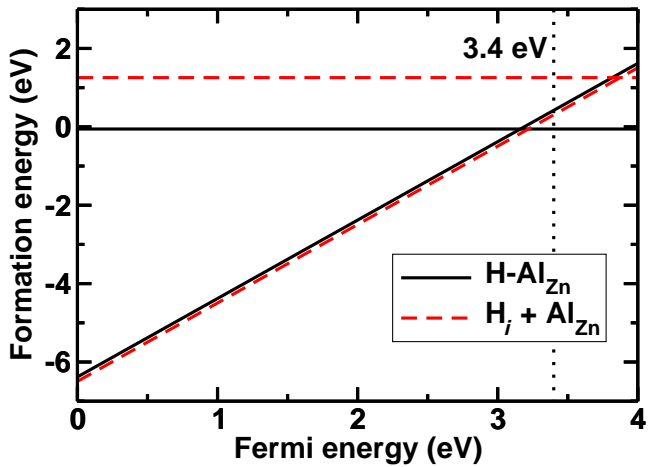}%
\caption{Formation energy as function of Fermi energy for the $\mathrm{H}-\mathrm{Al}_{\mathrm{Zn}}$ complex (full black curve) and for two noninteracting interstitial hydrogen $\mathrm{H}_i$ and substitutional $\mathrm{Al}_{\mathrm{Zn}}$ impurities (dashed red curve), in the oxygen poor limit.}\label{figAl}
\end{figure}

\begin{figure}
\includegraphics[width=8.5cm]{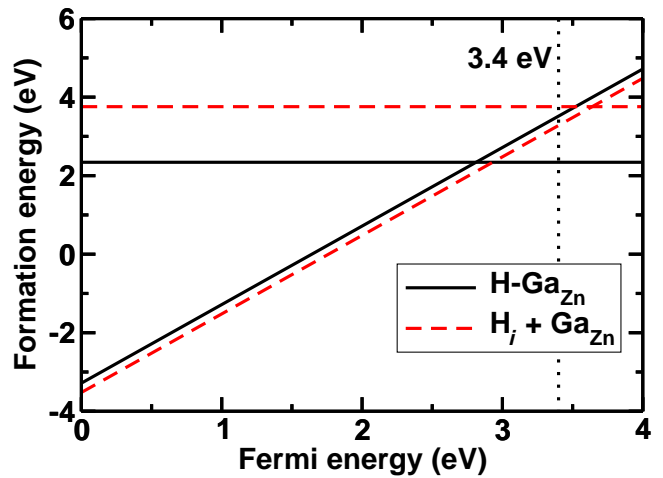}%
\caption{As Fig.~\ref{figAl}, but for Ga-H co-doped ZnO.}\label{figGa}
\end{figure}

\begin{figure}
\includegraphics[width=8.5cm]{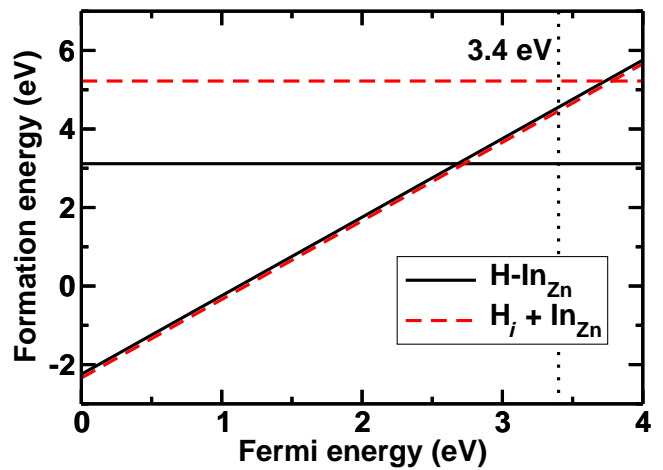}%
\caption{As Fig.~\ref{figAl}, but for In-H co-doped ZnO.}\label{figIn}
\end{figure}

For all charge states, we find that the lowest formation energy is obtained when the H atom is located close to the octahedral interstitial position next to an Al, Ga or In atom. This formation energy is compared with the formation energy for two non-interacting impurities, obtained by adding the formation energy of a single interstitial H atom ($\mathrm{H}_i$) and a single Al, Ga or In dopant in ZnO ($\mathrm{X}_{\mathrm{Zn}}$). The results are shown in Fig.~\ref{figAl}, Fig.~\ref{figGa} and Fig.~\ref{figIn} for the co-doping cases H-Al, H-Ga and H-In, respectively. The shown curves are for the oxygen poor limit. For the oxygen rich limit, all curves shift down with 3.43 eV. This choice has no influence on our conclusions, as they are only based on the crossing points between the formation energy curves for different charge states. As both H and (Al, Ga or In) can give away a single electron, the $+2$ charge state is the ground state for small values of the Fermi energy. (The $+1$ case is not shown, as it never becomes the ground state.) It is clear that the case of two non-interacting, thus infinitely separated, impurities has the lowest formation energy. In this $+2$ charge state, both shallow donors repel each other. More importantly, however, for Fermi energy values close to the band gap of 3.4 eV, the neutral H-(Al,Ga,In) complex has the lowest formation energy. Furthermore, the crossing between the non-interacting charge case $+2$ and the neutral complex occurs in the band gap, i.e. a deep level is formed. This crossing occurs at 0.18 eV, 0.47 eV and 0.67 eV below the bottom of the conduction band for Al-H, Ga-H, and In-H, respectively.  This is a consequence of the large decrease in formation energy of the neutral complex in comparison to the formation energy of two non-interacting neutral impurities: 1.31 eV for Al-H, 1.42 eV for Ga-H, and 2.11 eV for In-H. As a consequence, the shallow donor H binds to and passivates the shallow donor (Al, Ga or In).

\begin{figure}
\includegraphics[width=8.5cm]{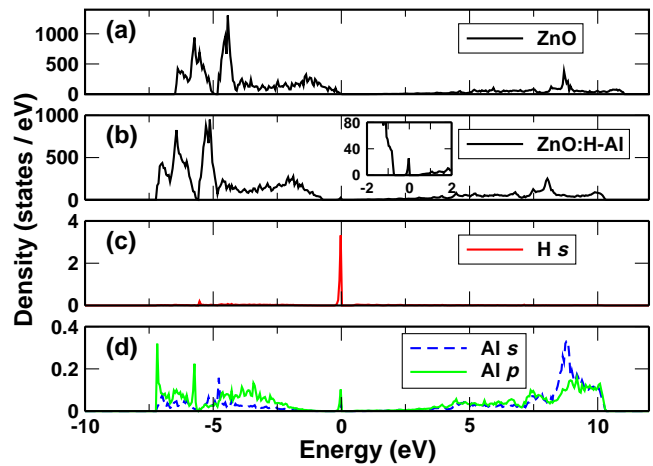}%
\caption{Calculated DOS, with zero energy taken at the Fermi level: (a) total DOS for the pure ZnO crystal; (b) total DOS for the Al-H co-doped ZnO, the inset is a zoom around the Fermi level; (c) projected (local) DOS on the H site; (d) projected (local) DOS on Al site.}\label{figDOS}
\end{figure}

The formation of a deep level is also clear from the density of states (DOS). As an example, the DOS for the Al-H co-doped case obtained within the LDA approximation is shown in Fig.~\ref{figDOS}. The neutral complex is considered in its ground state configuration. Comparing Figs.~\ref{figDOS}(a) and (b) for the pure ZnO crystal and the co-doped case, a shark peak appears in the ZnO band gap. From the projected (local) DOS plots on H (Fig.~\ref{figDOS}(c)) and Al (Fig.~\ref{figDOS}(d)) we observe that this deep donor level has H $s$ and Al $p$ character. While an individual interstitial $\mathrm{H_i}$ and a substitutional Al$_{\mathrm{Zn}}$ atom in ZnO have both a defect level in the conduction band with these characters, a bonding level is formed in the case they form a complex, which lowers the energy to a value in the band gap.

\begin{figure}
\includegraphics[width=8.5cm]{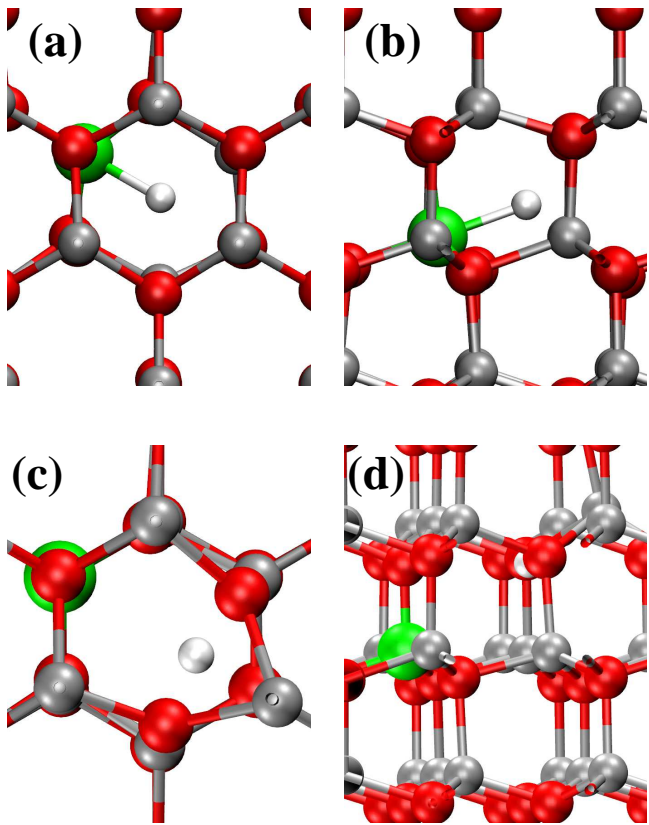}%
\caption{(a) and (b) The Al-H neutral complex. (c) and (d) The $+2$ charge case of the Al-H complex. Red atom is Zn, grey atom is O, green atom is Al, white atom is H. These figures are generated by VMD version 1.8.7~\cite{vmd}.\label{config}}
\end{figure}

To gain a better understanding of the physics behind this finding, we considered the bond length in the neutral complex. The bond between both shallow donors is clearly reflected in the interatomic distances. In the neutral case, the equilibrium distance between the interstitial H atom and the substitutional group III atom is 1.86 \AA\ for the Al case, 1.77 \AA\ for the Ga case and 1.87 \AA\ for the In case. However, in the $+2$ case, these distances increase till close to 3.15 \AA\ for all three cases, showing a clear repulsion. As an example, for the Al-H co-doped ZnO, the structure of the neutral Al-H complex is shown in Fig.~\ref{config}(a) and (b), and for the repelling $+2$ case in Fig.~\ref{config}(c) and (d). It is also interesting to compare these bond lengths in the case of the neutral complex with those of the corresponding molecular species AlH, GaH and InH, which are 1.65 \AA, 1.66 \AA, and 1.84 \AA~\cite{aldridge2001}. The Al-H and Ga-H bond length are thus longer in comparison to the length in the molecular species, especially in the Al-H case, while the In-H bond length is much closer to the molecular structure InH. This shows that there is indeed a bond in the neutral H-X complex, although weaker than in the molecular case.

To support this view and to propose an experimental way to detect these complexes, we also studied a vibrational property. The microscopic geometry for hydrogen in semiconductors is usually determined through observation of the infrared spectrum of the related local vibrational mode. In order to facilitate the experimental observation of the H-(Al, Ga, In) complex, we calculate the H-X stretching mode. As the H atom is so light we also take into account the anharmonic effect. Therefore the potential curve is fitted by the following formula~\cite{limpijumnong03}
\begin{eqnarray}
 V(x) = \frac{k}{2}x^2 + \alpha x^3 + \beta x^4. \label{eq:V(x)}
\end{eqnarray}
We have taken displacements up to $\pm$ 30 \% of the X-H bond length along the bond direction.
By applying perturbation theory to the one-dimensional Schr\"odinger equation an approximated analytical
solution in the case of the anharmonic potential [eq.~(\ref{eq:V(x)})] is given by~\cite{landau81}
\begin{eqnarray}
 \omega = \omega_{\mathrm{H}} + \Delta \omega_{\mathrm{A}} = \sqrt{\frac{k}{\mu}} - 3 \frac{\hbar}{\mu}\left[ \frac{5}{2}\left( \frac{\alpha}{k}\right)^2- \frac{\beta}{k}\right],
\end{eqnarray}
where $\omega$ is the frequency of the mode, $\omega_H$ and $\Delta\omega_A$ its harmonic and anharmonic contribution, and $\mu$ is the reduced mass. The obtained results for the frequency of the local vibrational modes are shown in Table~\ref{tab:freq}. Again we can compare with the corresponding frequencies of the molecular species. These results are also included in Table~\ref{tab:freq} under $\omega_{mol}$ (taken from Ref.~\cite{aldridge2001}). Note that the obtained frequencies for the complex are smaller than in their molecular equivalents. The difference is most pronounced for Al-H, and the smallest for In-H. The larger the difference, the weaker the bond is in the complex in comparison with the molecular equivalent. This trend is consistent with the trend that was observed for the bond lengths. This then also explains why the defect level of the complex is deepest in the In-H case.

To summarize, we have studied by first principles calculations the co-doping of ZnO with hydrogen and a group III element (Al, Ga, or In). While these impurities are known to be shallow donors in ZnO, and are therefore expected to repel each other, we show on the contrary that they attract each other, resulting in a neutral complex which forms a deep level. Hydrogen thus passivates the group III impurity in ZnO, which puts a limit on its $n$-type conductivity. In semiconductors in which hydrogen shows amphoteric behavior, passivation of a shallow impurity does not mean that a complex is formed, the formation of such a complex may just be a consequence of the passivation. In ZnO however, in which hydrogen acts as a shallow donor, it is the pairing between hydrogen and the substitutional shallow donor into a neutral complex that realizes the passivation.

\begin{acknowledgments}
We gratefully acknowledge financial support from the IWT-Vlaanderen through the ISIMADE project, the FWO-Vlaanderen through project G.0191.08, and the BOF-NOI of the University of Antwerp. This work was carried out using the HPC infrastructure of the University of Antwerp (CalcUA) a division of the Flemish Supercomputer Center VSC, which is funded by the Hercules foundation. M. M. acknowledges financial support from the GOA project ``XANES meets ELNES'' of the University of Antwerp.
\end{acknowledgments}


\begin{table}
\caption{Calculated harmonic frequencies ($\omega_H$), anharmonic contributions ($\omega_A$) and total frequencies ($\omega$) of the local vibrational mode of the X-H complex, and the corresponding frequency ($\omega_{mol}$) in the corresponding molecular structure, taken from Ref.~\cite{aldridge2001}. All frequencies are expressed in cm$^{-1}$.}\label{tab:freq}
\begin{ruledtabular}
\begin{tabular}{ccccc}
Complex & $\omega_H$ & $\Delta\omega_A$ & $\omega$ & $ \omega_{mol}$~\cite{aldridge2001} \\  \hline
H-Al & 1266 & -54 & 1212 & 1682\\
H-Ga & 1265 & -64 & 1201 & 1604\\
H-In & 1350 & -71 & 1280 & 1475\\
\end{tabular}
\end{ruledtabular}
\end{table}





\providecommand{\noopsort}[1]{}\providecommand{\singleletter}[1]{#1}%
\end{document}